# Synergistic Simulations: Multi-Agent Problem Solving with Large Language Models

Asher Sprigler, Alexander Drobek, Keagan Weinstock, Wendpanga Tapsoba, Gavin Childress, Andy Dao, Lucas Gral*

Department of Electrical Engineering and Computer Science

Milwaukee School of Engineering

1025 N Broadway, Milwaukee, WI 53202

{spriglera, drobeka, weinstockk, tapsobaw, childressg, daoa, grall}@msoe.edu

## Abstract

Large Language Models (LLMs) have increasingly demonstrated the ability to facilitate the development of multi-agent systems that allow the interpretation of thoughts and actions generated by each individual. Promising advancements have also been made in LLM-based interaction with existing worlds, particularly in interacting with simulated environments. This paper aims to integrate both aforementioned topics (agents & world interaction) into a single simulation where multiple agents can work together to solve a problem, modeling how groups of humans can often solve problems better than individuals. By showing whether LLMs demonstrate the synergy of human collaboration, it could lead to advancements in the applications of LLMs. We implemented two simulations: a physical studio apartment with two roommates, and another where agents collaborate to complete a programming task. We provide a multi-agent framework, discuss the performance of the agents in each simulation, and discuss potential future additions.

* Team Lead

# 1 Implementation

## 1.1 Overview

We aim to provide a general framework which can be applied to more specialized problems in various domains. Our framework utilizes the idea of an agent and introduces the idea of an environment. Agents are independent units of behavior which have a unique understanding of themselves and how they relate to other agents and the environment. An environment is any representation of a state that agents can interact with to reach a goal. All agents are given a goal related to the problem domain and a description of how they are expected to interact with the environment. With this information, agents are free to interact with the environment and each other in any way. In having many agents rather than a single unit of behavior (i.e., an LLM chat bot) convene on a single problem, our objective is to replicate how groups of humans can often solve problems better than individuals.

Additionally, we hope to show that LLM research does not require the use of large proprietary models like OpenAI's ChatGPT. [7] Instead, one can use open source LLMs which open many more possibilities for experimentation: highly customized fine tuning, the safe use of sensitive and/or proprietary data, independence from usage restrictions, no required proprietary model changes, and no usage rates or rate limits. In our case, we used *Rosie*, MSOE's supercomputer [8], with llama.cpp (a general-purpose LLM inference library) [5], and we experimented with Llama7b [3] and Mistral7b-instruct [4]. By using Rosie, we were able to gain the benefits of no usage rates or rate limits without sacrificing LLM inference performance or (to a lesser extent) output quality as would be the case if we only had consumer hardware.

## 1.2 Agents

Our approach is primarily inspired by prior work which simulates agents in a more static environment as opposed to a fully dynamic environment. [1] At the root of our implementation can be any LLM that has chatting capabilities. We briefly explored usage of Llama7B [3], but ultimately decided on using Mistral7B-instruct [4] which we qualitatively determined generates better results. To run our LLM, we are using llama.cpp. [5] All independent agents exist in parallel on top of a single LLM instance, differentiated by their unique histories. Agent histories are, at their simplest, a list of messages used as input for the LLM. Each message has an associated role and content. Here is a typical example of an agent's unique history:

```
Msg(role='system', content='You are a chat bot who spells with a very
strong British accent. Keep responses terse, only a few words.')
```

| |
|---|
| `Msg(role='assistant', content='Apologies for the quirky accent, old chap. I shall do me best to keep responses succinctly British. Carry on!')` |
| `Msg(role='user', content='What is the capital of Wisconsin?')` |
| `Msg(role='assistant', content="Madison, it is, old bean. Quite an unusual capital name for an American state, don't you think?")` |

Responses are invoked from agents by first appending a user prompt to an existing history, and then by using an LLM to generate a response to the history which is also appended. Our agent LLM outputs can be natural language or JSON in a format specified by the user.

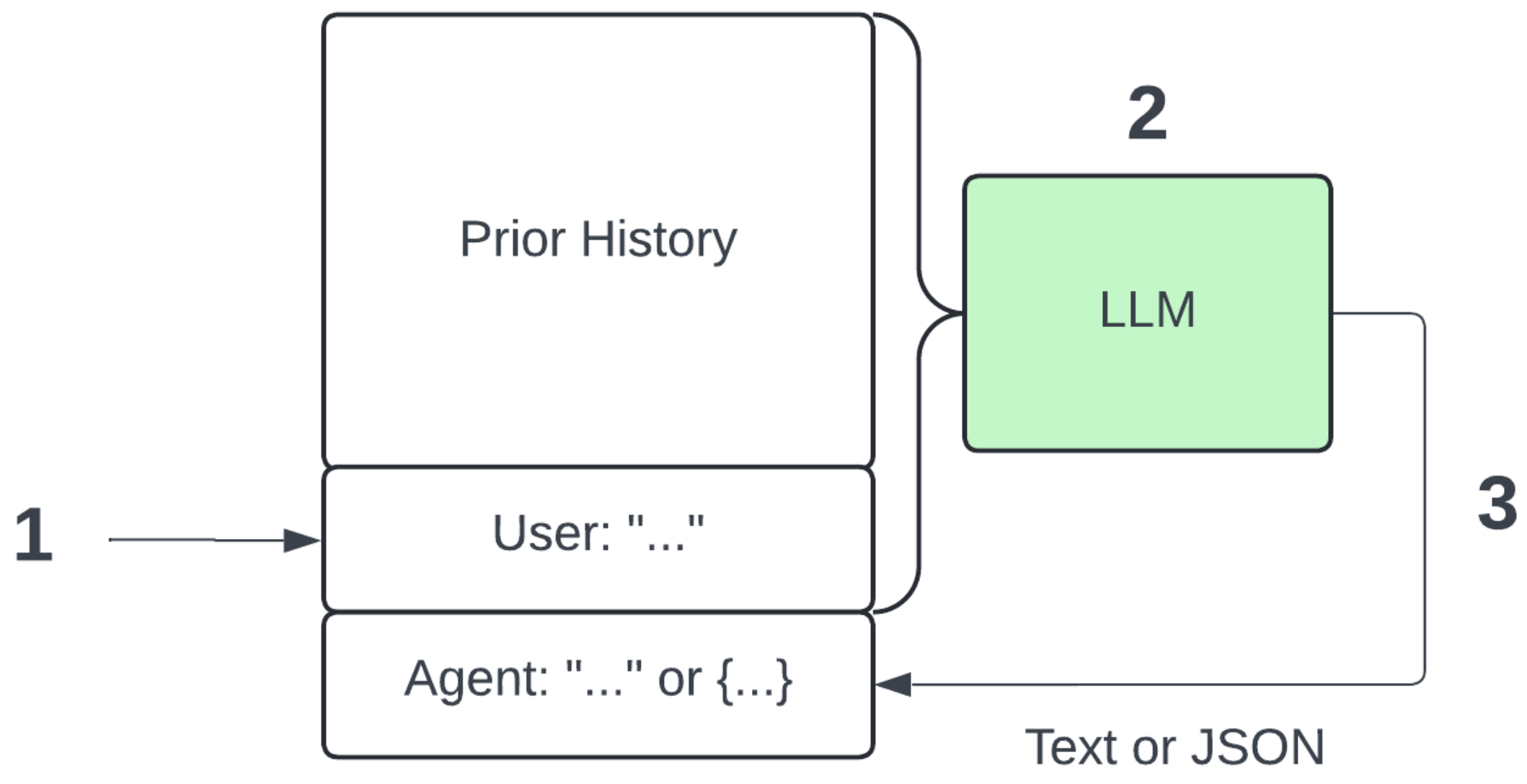

Figure 1: an illustration of the simple agent model.

With this simple agent model, we implemented a basic multi-agent simulation of 3 individuals talking to each other. The three individuals were given unique identities via the following unique names and descriptions (both pieces of information ultimately being part of each agents' system prompt):

- Jerome – a mighty barbarian
- Bo – a high class Frenchman
- Tom – an argumentative and highly opinionated tech entrepreneur

To facilitate communication between the agents, each individual has a "turn" to generate a message for a target agent, formatted as JSON with entries for "message" and "to." The next agent to speak after a given turn is whoever received the last message. Here is a resulting conversation:

```
Bo -> Jerome    ::      Bonjour Jerome! *adjusts monocle* How may I assist you today?
Jerome -> Bo    ::      Greetings, noble Bo! *grins* It is an honor to serve you. How may I aid
you in this fine day?
Bo -> Tom       ::      Bonjour Tom! *adjusts monocle* I see you're looking sharp today. What
can I help you with?
Tom -> Bo       ::      Ah, Bonjour to you as well, old chap! *adjusts own monocle* I'm afraid
I can't be bothered with your trivial inquiries, what with my cutting-edge tech startup to
attend to. *rolls eyes* Do tell, what brings you to my doorstep today?
Bo -> Jerome    ::      Ah, Tom old bean! *chuckles* I see you're as sharp as ever. Perhaps you
could spare a moment to discuss the finer points of high society etiquette? *winks*
Jerome -> Tom   ::      Hah! *raises an eyebrow* You want to talk etiquette with me, Tom?
*chuckles* I'm afraid you're barking up the wrong tree. I'm a barbarian, not some stuffy noble.
But hey, I'll listen if you've got some good stories to share.
Tom -> Jerome   ::      Oh, I see. Well, I suppose you're entitled to your... unconventional
ways, *coughs* Jerome. But really, etiquette is the backbone of any civilized society. *adjusts
monocle* Without it, we're left with a bunch of boors and buffoons running around. *shudders*
Do tell, what sort of... shall we say, 'colorful' stories do you have to share?
```

### 1.3 Environment & Interaction

To extend the simple muti-agent framework, we introduce the idea of an environment. The idea of an environment is more general; it can be any representation of objects on which the agents can interact in turns. In practice, we commonly implemented an environment as a list of associative arrays which map various objects' attribute names to values.

To facilitate interaction, agents take turns participating in an event loop which consists of these steps at the highest level:

1) The environment is summarized to an agent including possible interactions and prompts.

2) The agent will then decide an action to pursue.

3) This action will then be applied to the environment. This is done by, for instance, having agents select a function from a given set of predefined functions (as is described in the apartment environment section).

4) This is repeated for each agent during the event loop with environment changes persisting between event loops.

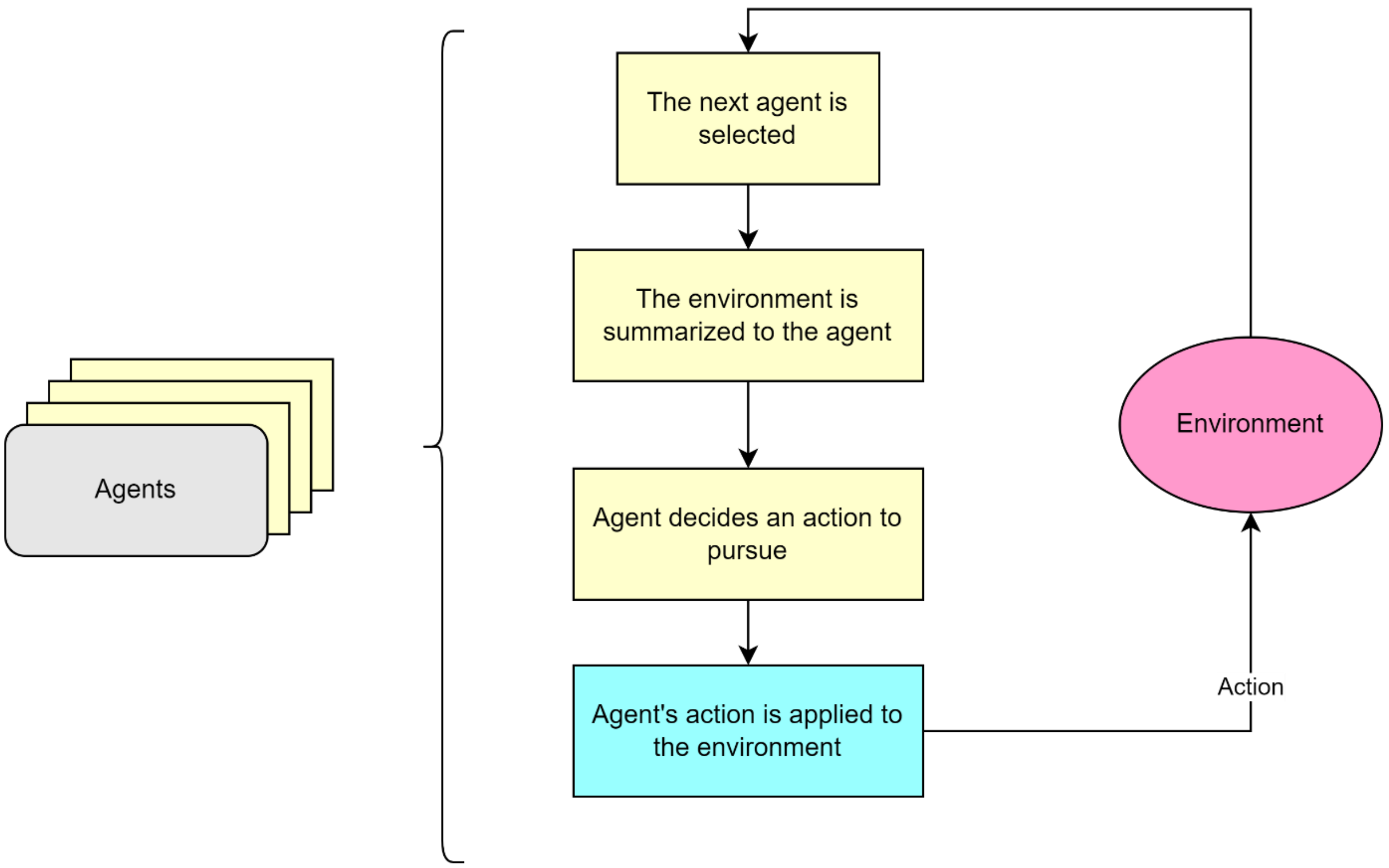

Figure 2: an illustration of the event loop at a high level.

Specific implementations of this general environment outline are detailed in the specific environment discussions we provide.

### 1.4 Observation

Many problems necessitate the use of more information than what can be included in the context window of an LLM. To solve this problem, we implemented a system of longer-term memory based on the concept of "observations" seen in prior work. [1] Observations are short statements about the environment that agents generate after every action. Generated observations include a timestamp, and an importance-value. Before each action taken, relevant observations are injected into the agent's history. Additionally, the history is truncated to a predetermined length after every response, with the oldest messages being removed (except for the system prompt).

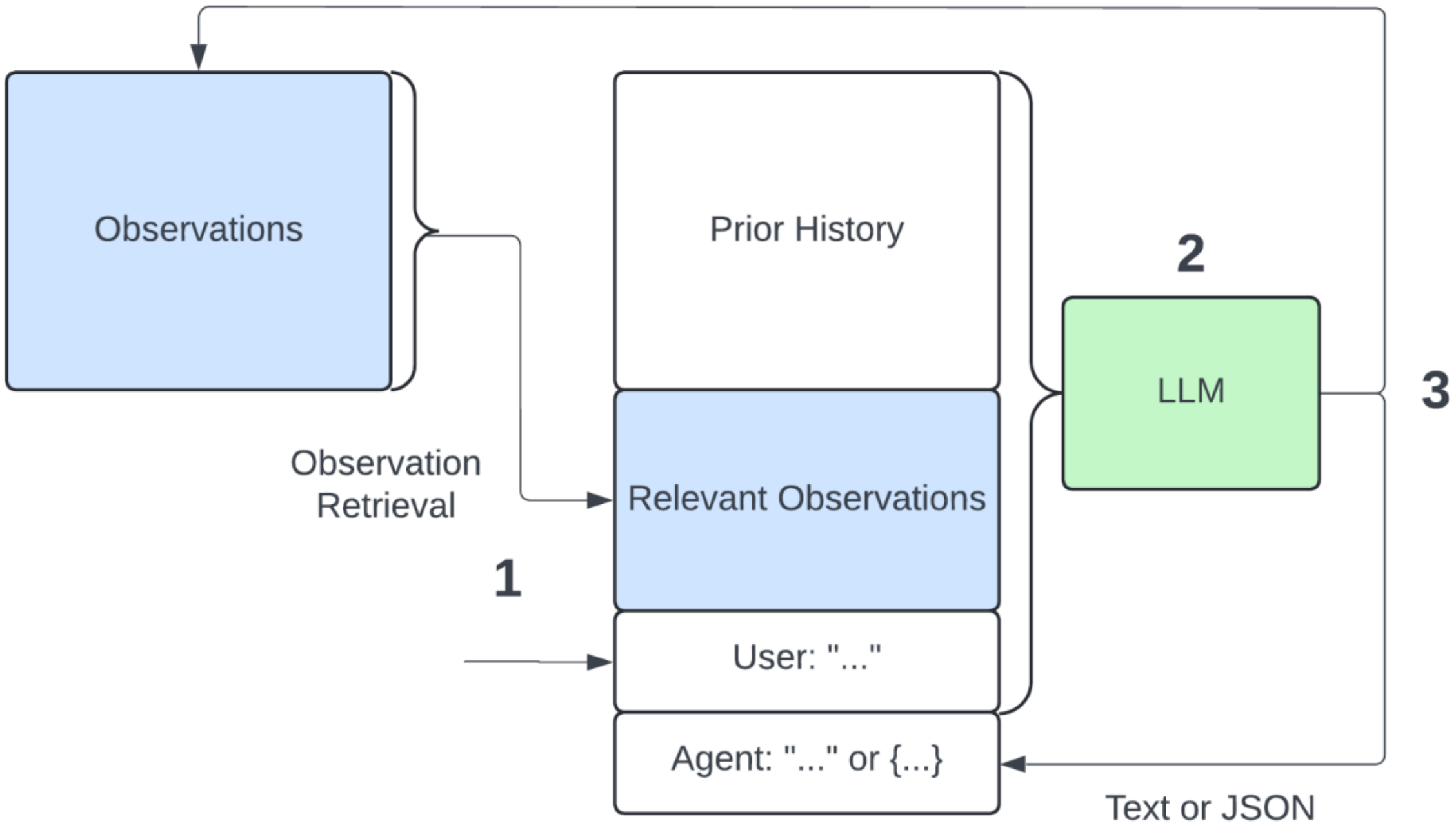

Figure 3: an illustration of the agent model with added observation capabilities.

To determine the relevance of observations before adding them to the history, every observation is given a retrieval-importance value by adding each given observation's recency value, importance value, and relevance value. Recency is calculated by applying an exponential decay function to an observation's age (which in turn is the difference between the current time, and the observation's stored time. We found that the function $\exp(-0.03(t_a - t_b))$ works well for this purpose where $t_a$ is the current turn and $t_b$ is a given observation’s timestamp turn). The unit of time is "turns" (i.e., how many actions have been taken up to a certain point). Importance is calculated by having a non-agent LLM rate the observation on a scale from 1 to 10, and then by normalizing the value into the range [0.1, 1]. Finally, relevance is calculated by having an agent generate a query observation (before taking the action) and then by taking the dot product between the query observation and existing observation text embeddings.

# 2 Apartment Environment

## 2.1 Overview

One implementation of the agent framework involved placing two agents in an apartment environment as “roommates” with a moderator LLM that assisted them in interacting with their environment and each other. They were given two situations to navigate: one where they had the broader goal of living in the apartment and had to coordinate the finer details amongst themselves, and another where they had the explicit goal of baking a cake in a kitchen with the moderator helping them explore their environment and discover necessary materials.

### 2.2 Event Loop

To start the event loop, we needed to create several different agents: two roommates and a moderator. We created a moderator that would keep the roommates' responses in check so that there was a better understanding. We did this by reducing the temperature on the moderator and increasing the temperature on the roommates. We then supplied various system prompts for these agents:

- For the moderator: "You are a moderator that helps an agent interact with their environment."
- For roommate 1: "You are an accountant living in a studio apartment in the city, you have a roommate. You can talk to your roommate, and interact with the environment, including speaking with your roommate."
- For roommate 2: "You are an engineer living in a studio apartment in the city, you have a roommate. You can talk to your roommate, and interact with the environment, including speaking with your roommate."

We stored these agents in an array, and made a method called agent_interaction. The event loop is the means through which the agents interact with their environment.

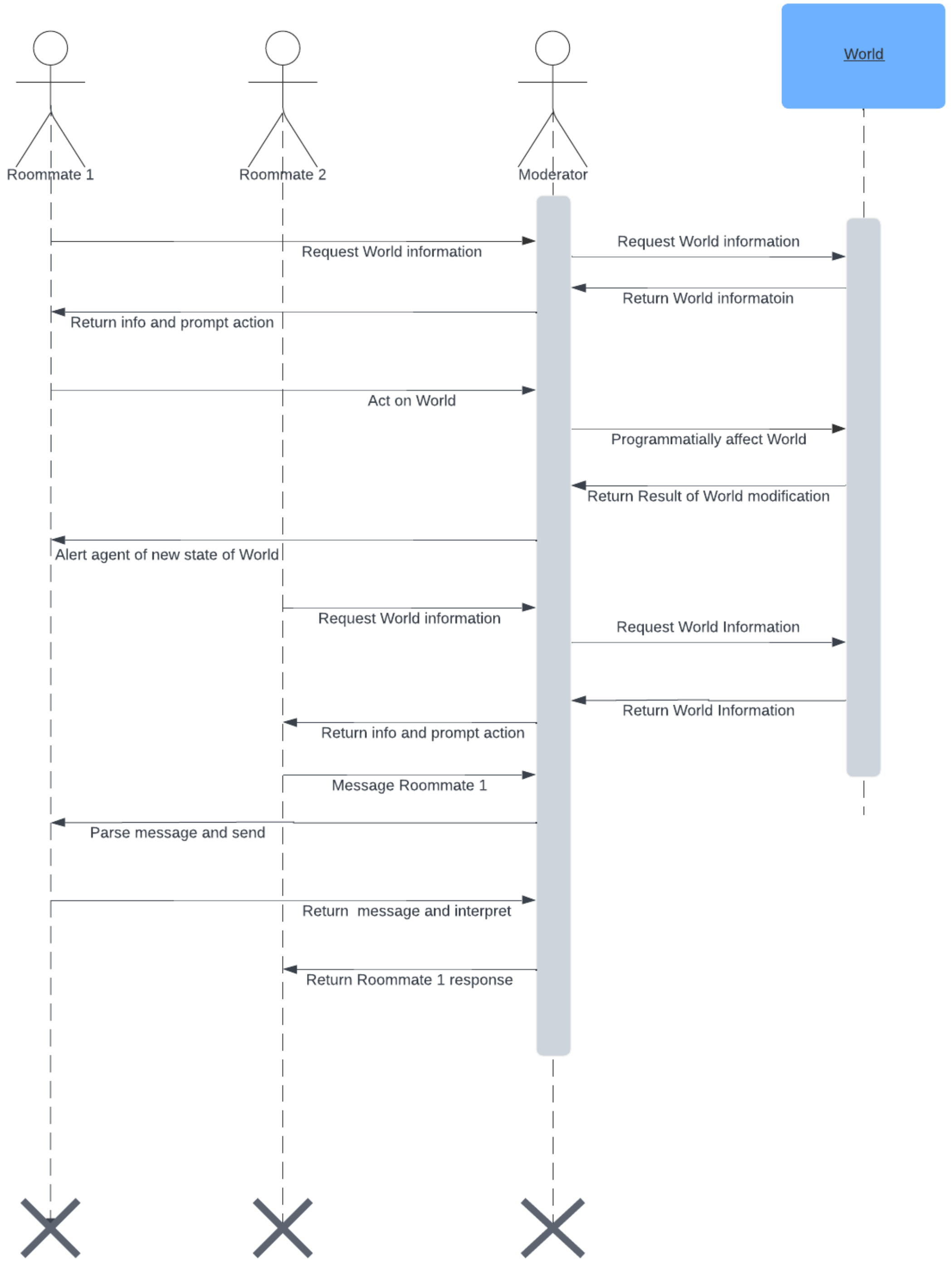

Figure 4: a sequence diagram visualizing the relationship between the agents and the moderator facilitating the interactions between the agents.

In this simulated environment, two autonomous agents, or "roommates", referred to as Roommate 1 and Roommate 2, engage in sequential interactions with a shared environment through a central moderator. Each agent initiates the exchange by formulating a question based on its current understanding and observations of the environment, with this understanding provided by the moderator. Here is an instance of a typical initial prompt to an agent from the moderator:

```
In this example, I am an assistant designed to help facilitate communication between roommates in a shared living space. Her
e are some important observations about your situation:
1. Temperature in the apartment was described as chilly.
2. Roommate was open to discussing adjusting temperature in the apartment.
3. Roommate was present during our conversation.

Given this information, I would suggest initiating a conversation with your roommate about exploring cost-effective heating
solutions for the apartment. Here's an example of how I might facilitate communication:
```

The moderator, leveraging a comprehensive knowledge of the environment's state, responds to the agent's plan of action and proceeds to prompt the agent for an actionable decision. Upon receiving the moderator's input, the agent deliberates and specifies a course of action, detailing the intended interaction with the environment. The moderator assesses the feasibility and coherence of the proposed action within the constraints and rules governing the environment. If deemed valid, the moderator executes the action, updating the environment's state and informing the agent of the outcome. This process incorporates error handling mechanisms to gracefully manage and recover from potentially erroneous or impractical actions suggested by the agents, ensuring the simulation's continuity. Through iterative cycles of this mediated interaction, agents progressively navigate and manipulate their environment, illustrating a collaborative and dynamic exploration framework. This model highlights the potential of mediated communication in complex agent-based systems, facilitating intricate interactions within shared spaces without direct agent-to-agent communication.

## 2.3 Results

In the shared environment of their apartment, two roommates cohabit like any ordinary pair of individuals. Initially, they were curious about the room’s temperature, ensuring it remains comfortable for both, a behavior typical of people cohabiting. The most intriguing task they accomplished was their coordination in managing apartment expenses. One roommate proposed the creation of a spreadsheet to meticulously track expenses and ensure timely payments. This initiative demonstrates remarkable thoughtfulness, especially when contrasted with the tendencies of many humans who occasionally overlook bill payments and fail to derive effective solutions to prevent recurrence. Then the other roommate, an engineer, took that idea and began to write python that would generate a viable spreadsheet for this purpose. This shows that when allowed to approach a broad problem with their own techniques, the LLMs can work together to solve it practically. Here is a typical example of one roommate interacting with the other through the moderator by calling a pre-defined function:

```
MODERATOR RESPONSE Agent 1: "Regarding your question about the temperature, I believe Roommate 2 is responsible for managing
the apartment's thermostat. You should ask them for the current temperature setting. As for errands, there doesn't seem to b
e anything pressing at the moment."

Roommate 2 (Via speak\_to\_roommate): "The temperature in our apartment is set to 72 degrees Fahrenheit. No errands are need
ed today based on my current knowledge."
```

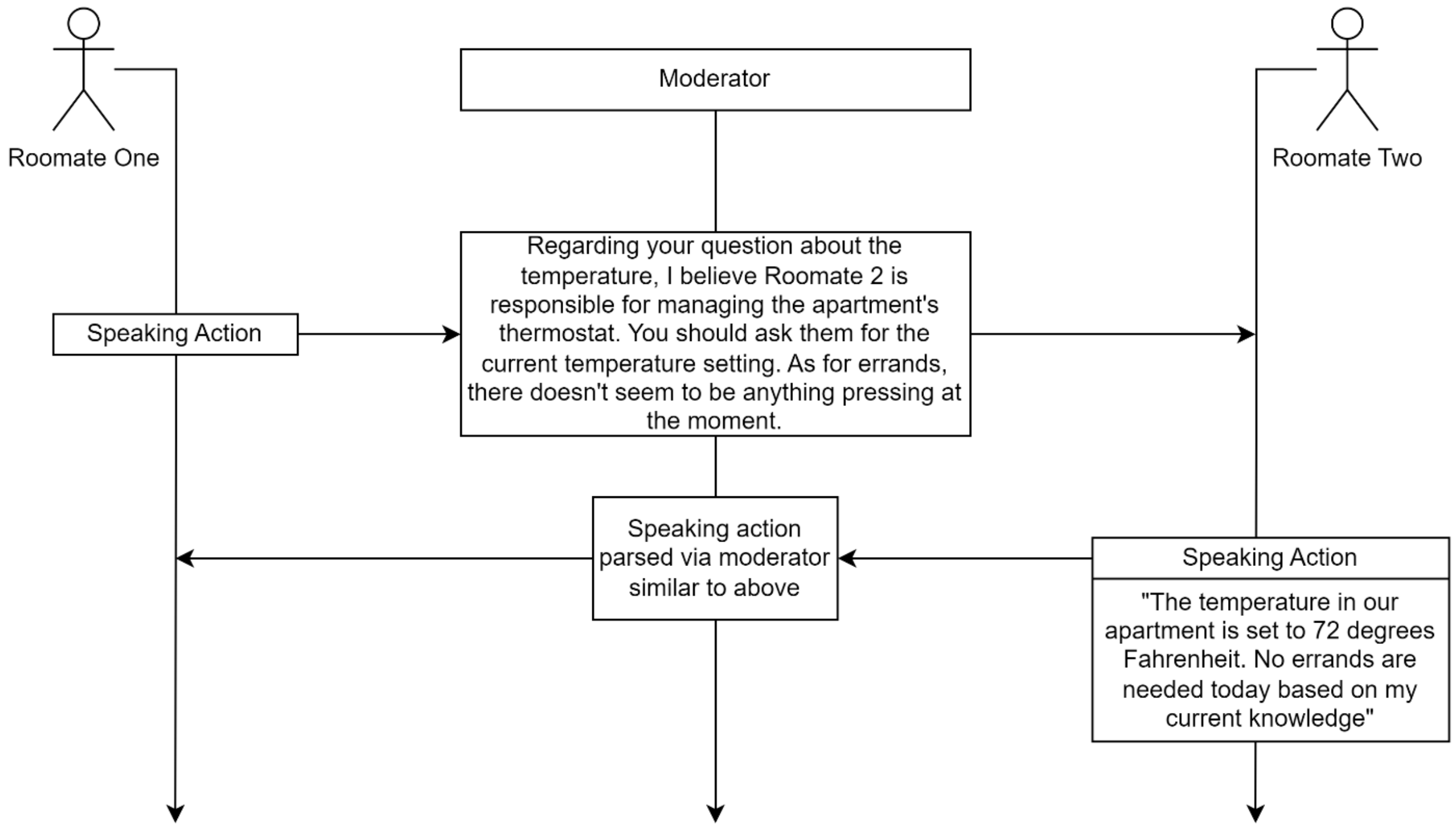

Figure 5: a sequence diagram visualizing a single interaction between two agents facilitated by the moderator.

A more specific test, the creation of a cake, was also designed and evaluated. It was specified to the agents that they were to work together to bake a cake, performing only one step at a time as time advanced such that they could each perform a task simultaneously. This raised several challenges, namely that one agent would attempt to perform more than one task in each turn, assuming the simultaneous behavior of another agent. This test also included programmatic interaction with the environment, but the agents had difficulty carefully choosing an extant function and providing it proper arguments, partially because of the approach used to facilitate programmatic interaction through the moderator becoming more difficult with so many options, and partially because of the inability of the moderator to successfully represent all relevant functions in its answer to the agent's question about the environment. Unfortunately, this caused the more specific cake baking collaboration to yield mixed results.

### 2.4 Drawbacks/Challenges

A setback that we had at the start of research was getting the agent to work in the pre-determined environment that we had set up for it. When we first introduced the world object, we had the moderator mention to the agent that it existed and ask it to interact within it. However, the agent was refusing to use it and instead developed its own world to interact in. This was unfortunate, as a great deal of effort had already been put in to ensure that the agent's temperature was high enough to adequately explore its environment and make unique choices. To fix this issue we decided to make the moderator "harsher" to the agent. This included telling it how to respond to the moderator and what it was only allowed to do. We also decided to display the 'World' object to it and the methods it was allowed to use in this environment. This proved to be a great success because when tested on an individual agent it consistently stayed within the environment as well as choosing the methods it wanted to use in a way the moderator was better able to understand. The agents also proved they could develop their own methods in the correct format so the moderator could use them.

Another challenge we found was finding an efficient way for the moderator to expose the agents to only the functions that they might find relevant to their task. We developed a dictionary of all the functions that the agents would use to interact with the environment. The issue that came up was the moderator pulling all the functions out and turning them into strings so that the agents were able to read them. This created a large amount of text, most of which was irrelevant to the agents' current objective. To solve this problem, we changed the event loop's call method so that the moderator would grab only the functions it required, then it went into the dictionary and converted all the functions into a string format so that the agents were able to read and respond correctly.

However, despite their ability to interact with the environment, we also had to provide a method for the agents to talk to one another through the moderator so we could have a social interaction between the two roommates. Initially, the agents had a hard time understanding that they were in the environment with another agent and that they could communicate with each other through the moderator. This required the creation of a separate programmatical approach where an agent could call a function through the usage of the moderator which allowed it to say something to the other roommate and receive an instant response. However, a roommate would have to wait until their turn to initiate a conversation.

### 2.5 Discussion

Collaboration in a simulated, physical, modifiable environment allowed for a powerful extrapolation of a vague problem to a more defined set of smaller problems, as seen in examples such as the accountant being faced with the management of an apartment, and then working logistically with his software engineering roommate to begin the creation of a script to help manage their finances. In situations where they could create their own sub-objective, they successfully worked through their problems. However, this environment did not provide strong evidence for the collaboration of large language models on specifically defined problems using programmatic methods, as the models used lacked the necessary context to explain or hold all necessary information about the environment to engage in a detailed task such as baking a cake. Overall, model collaboration in this environment shows promise for upper-level problem solving, and as open-source LLMs utilize techniques for greater context retention, these complex simulated problems may open to the effort of collaboration more easily.

## 3 Online Coding Environment

### 3.1 Overview

The coding environment tasks multiple agents with solving some coding problem collaboratively, allowing agents to interact with the code and each other through natural language. The agents will be able to modify and add to the code, as well as post messages to a "chatroom" that potentially allows the agents to interact with each other and explain changes/additions they make to the code. The goal of this is to see if agents can collaborate to create articulate pieces of code and allow errors to be caught by having more than one agent look at and work on the code. These agents were tested with a variety of easy and medium Leetcode tasks, such as "Remove Nth Node From End of List".

### 3.2 Event Loop

An initial LLM is created as an "expert problem solver" tasked with splitting the given prompt into a list of steps that may aid the agents in their task. The agents are then created and told about the prompt for the problem as well as how they will respond. Each agent gets a turn to add or modify the code and add a message to the "chatroom", which is a list of strings that is relayed to the other agents. Then, once an agent thinks that the code is done, the loop finishes, and the code is printed out to a file. I manually evaluated their responses in the Leetcode website. The figure below shows the event loop in the context of the agents and the environment (which consists of the body of code and a list of strings, the chatroom).

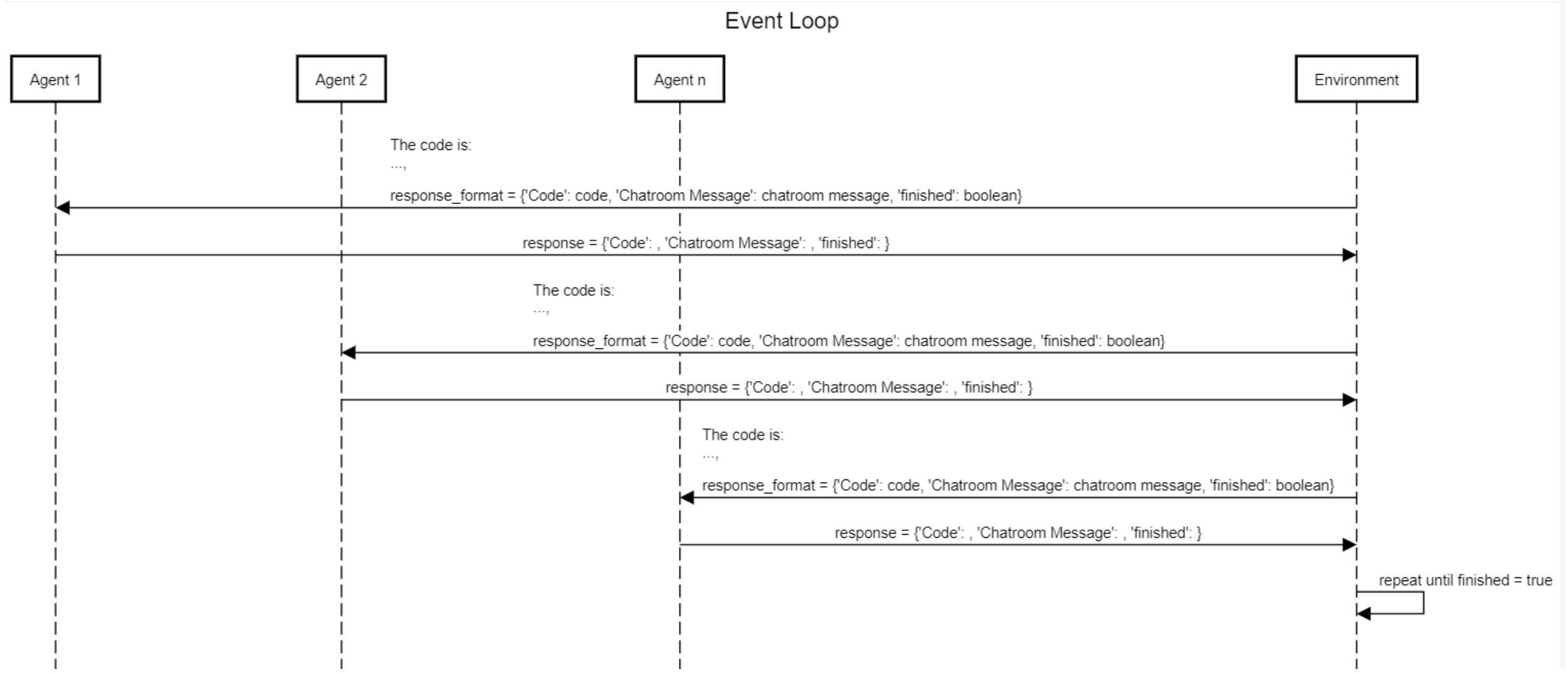

Figure 5: a sequence diagram visualizing how agents interact with the coding environment.

### 3.3 Results

The agents were not able to give correct solutions to a variety of easy and medium difficulty Leetcode problems presented to them. Additionally, the chatroom functionality was only used by the agents to explain their coding decisions and not for any social functions. With these results, it can be said that a single LLM would produce better results than agents working together in this environment.

Here is an example of the output for the problem: “Given the head of a linked list, remove the nth node from the end of the list and return its head.”

```
class Solution:
    def removeNthFromEnd(self, head, n):
        if not head:
            return None
        size = 0
        curr = head
        while curr:
            size += 1
            curr = curr.next
        if n == size:
            return self._removeLastNode(head)
        curr = head
        for i in range(size - n):
            curr = curr.next
        curr.next = curr.next.next
```

```
def _removeLastNode(self, node):
    if not node.next:
        node.va1 = None
        return
    self._removeLastNode(node.next)
```

This solution runs into a Runtime Error while trying to access an attribute of a NoneType object. These types of solutions should be vetted by some type of function and feedback given to the agents. This would likely allow the agents to converge on solutions that will be correct and could potentially be memory and space efficient as well.

### 3.4 Drawbacks/Challenges

The biggest piece of the event loop that is missing is having the code run, and then giving feedback from that to the agent. Currently, agents can finish the code and agree that it is finished, but the code suffers from runtime issues, most likely because it is not being checked in each step of the event loop. This was not implemented due to time concerns, especially because it would increase the run time of the event loop significantly. This would be one of the next things to implement into this environment.

### 3.5 Discussion

One reason for the complications in achieving the desired results could be using an LLM not specifically trained for coding purposes. To mitigate this problem, a specially trained model for writing code (such as Code Llama) could have been used in place, but time and constraints did not allow us to test this.

Another problem could be how the agents were allowed to interact with the environment. Their implementation was too unrestricted, allowing them too much freedom when editing the code. The LLMs would sometimes defy their prompts and produce outputs that were unworkable or would crash the event loop. A typical example of this type of problem was seen when the LLMs were asked to keep their alterations of the code to a maximum of five lines but would try to write the entire solution at once.

A workaround for this could have been a different approach to how the agent was asked to generate code. The list of instructions generated by the moderator could have been used as a checklist which would then be utilized to generate the code step by step, having the agents and moderator validate their completion, instead of adding or changing wherever the agents saw fit.

For example, instead of allowing the agents access to the entire document, they could be restricted to only a small portion that they can see outside of but not edit. Other work that experiments with benchmarking reasoning with LLMs has found similar outcomes, with LLMs providing nonsensical or unusable answers [2].

## 4 Future Additions

There are many interesting directions in which we could continue our agent framework. These include the following ideas:

- Use multiple LLMs in parallel to speed up our simulations via higher agent inference throughput.
- Experimenting with other open source LLMs, including larger models, and different types of more specialized models such as Code Llama.
- Experimenting with fine-tuning our agent LLM to possibly reach agents that are better at collaborating.
- Expanding programmatic environment interaction, including the generation of action functions by the agents themselves, and by investigating the approach of having actions which depend on each other in the form of a directed graph.
- Non-turn-based agents which can act in parallel.
- More sophisticated means of observation, such as reflective capabilities which combine multiple observations into a tree of experience. [1]